\documentclass{moriond}

\def\Journal#1#2#3#4{{#1} {\bf #2}, #3 (#4)}

\def\PRD{{\em Phys. Rev.} D}

\def\be{\begin{equation}}
\def\ee{\end{equation}}
\def\bea{\begin{eqnarray}}
\def\eea{\end{eqnarray}}

\newcommand{\Photo}{}

\usepackage{bm}
\usepackage{amsmath}
\newcommand{\dd}{\text{d}}
\newcommand{\Var}{\text{Var}}

\begin{document}
\vspace*{4cm}
\title{Scattering transforms on the sphere, application to large scale structure modelling}

\author{L. Mousset$^{1}$, E. Allys$^{1}$, M. A. Price$^{2}$, J. Aumont$^{3}$, J.M. Delouis$^{4}$, L. Montier$^{3}$, J. D. McEwen$^{2}$.}

\address{$^{1}$Laboratoire de Physique de l’\'Ecole Normale Supérieure, ENS, Univ. PSL, CNRS, Sorbonne Univ., Univ. Paris Cité, 75005 Paris, France\\    $^{2}$Mullard Space Science Laboratory, UCL, Holmbury St Mary, Dorking, Surrey RH5 6NT, UK\\
$^{3}$Institut de Recherches en Astrophysique et Planétologie, Univ. de Toulouse, CNRS, CNES, UPS\\
$^{4}$Laboratoire d’Océanographie Physique et Spatiale, Univ. Brest, CNRS, Ifremer, IRD, Brest, France}

\maketitle\abstracts{Scattering transforms are a new type of summary statistics recently developed for the study of highly non-Gaussian processes, which have been shown to be very promising for astrophysical studies. In particular, they allow one to build generative models of complex non-linear fields from a limited amount of data. In the context of upcoming cosmological surveys, the extension of these tools to spherical data is necessary. We develop scattering transforms on the sphere and focus on the construction of maximum-entropy generative models of astrophysical fields. The quality of the generative models, both statistically and visually, is very satisfying, which therefore open up a wide range of new applications for future cosmological studies.}

\section{Introduction}
\label{sec:intro}

Scattering transforms (ST) are a recently developed class of summary statistics for the study of non-Gaussian processes~\cite{Bruna2012}. They are inspired from neural networks but do not require any training step to be computed. Introduced recently in astrophysics~\cite{Allys2019}\,\cite{Allys2020}, ST have since demonstrated their ability to characterize highly non-Gaussian processes, for instance for parameter estimation and classification tasks.

Another feature of ST is that they allow one to build very efficient generative models of physical fields, in a maximum entropy framework~\cite{bruna2019multiscale}. This allows one to sample new approximate realisations of a given process relying only on its ST statistics, that can be estimated even from a small amount of data, sometime even a single example image~\cite{Allys2020}\,\cite{Cheng2023}\,\cite{regaldo2023generative}\,\cite{price2023fast}.

While these promising ST generative models have mainly been developed for 2D planar data, the adaptation of these tools to spherical data is necessary for cosmological analysis, especially for the next generation of full sky surveys such as LiteBIRD for the cosmic microwave background polarization, or Rubin-LSST and Euclid for study of the large scale structures of the Universe. The extension of ST to spherical data however raises some difficulties: the definition of a directional convolution with oriented filters~\cite{mcewen:s2let_spin}\,\cite{mcewen:s2let_localisation}, as well as the transposition of the planar translations which appear in certain ST representations. In this paper, we propose an adaptation of state-of-the-art ST to spherical fields. As a first step, we restrict ourselves to homogeneous fields with properties that do not depend on the position on the sphere. This naturally leads us to cosmological fields. In this proceeding, we only show the results for a weak lensing field from the Large Scale Structures (LSS) of the Universe using the CosmoGrid data set~\cite{Cosmogrid}. For this field, we construct and validate a ST generative model from one single example image. Generative models built from other fields, such as the thermal Sunyaev-Zeldovitch effect are presented in Mousset et al. (2024)~\cite{mousset2024}.

\section{Scattering covariance on the sphere}
\label{sec:scatcov} 

ST refer to a family of summary statistics which includes in particular the Wavelet Scattering Transforms (WST)~\cite{Allys2019} or the Wavelet Phase Harmonics (WPH)~\cite{Allys2020}. In this work, we have considered the scattering covariances (SC), or scattering spectra~\cite{Cheng2023}. We chose the SC, because they only rely on convolutions, and not on translations as the Wavelet Phase Harmonics~\cite{Allys2020}, which are difficult to univocally define on spherical maps.

SC statistics are computed from wavelet transforms, which are obtained by convolving an initial map with a set of wavelet filters, where each filter extracts the local information at a particular scale (labelled by $j$) and orientation (labelled by $\gamma$). Wavelet filters need to be localized both in pixel and harmonic space. 

To compute the wavelet transforms, various convolutions on the sphere can be considered. In this work we follow the standard directional convolution formalism presented in, e.g., McEwen et al. (2015)~\cite{mcewen:s2let_spin}. The directional convolution $I \star \Psi^j$ of a field $I$ with a wavelet $\Psi^j$ consists in applying a rotation by a set $\rho = (\alpha,\beta,\gamma)$ of Euler angles of the wavelet $\Psi^j$ initially located at the north pole, before computing an inner product between the wavelet and the field $I$:
\begin{equation}
    (I \star \Psi^j)(\rho) 
    = \int_\Omega  I(\bm\omega) [R_\rho \Psi^j (\bm\omega)]^*  \dd \Omega,
\end{equation}
where $R_\rho$ is the rotation by Euler angles $\rho$, and $*$ stands for complex conjugation. From $(I \star \Psi^j)(\rho)$, we can identify ($\beta, \alpha$) with the spherical coordinates $\bm\omega = (\theta, \varphi)$ and $\gamma$ to the orientation which is probed in the convolution. In this way, we obtain oriented wavelet coefficients, that we describe with the following shorthand notation
\begin{equation}
    (I \star \Psi^{j,\gamma})(\bm\omega) \equiv (I \star \Psi^j)(\alpha = \varphi, \beta=\theta, \gamma).
\end{equation}

SC statistics characterize the power and sparsity at each scale, as well as interaction between different scales. They are built from successive applications of wavelet transforms and modulus operators, followed by average and covariance computations~\cite{Cheng2023}. We consider two coefficients at a single oriented scale $\lambda_1 = (j_1, \gamma_1)$:
\begin{equation}
    S_1^{\lambda_1} = \langle |I \star \Psi^{\lambda_1}| \rangle \, , \quad
    S_2^{\lambda_1} = \langle |I \star \Psi^{\lambda_1}|^2 \rangle \, ,
    \label{eq:P00}
\end{equation}
and two coefficients that characterize the couplings between two and three oriented scales:
\begin{equation}
    S_3^{\lambda_1, \lambda_2} = \text{Cov} \left[I \star \Psi^{\lambda_1}, |I \star \Psi^{\lambda_2}| \star \Psi^{\lambda_1}\right], \quad
    S_4^{\lambda_1, \lambda_2, \lambda_3} = \text{Cov} \left[|I \star \Psi^{\lambda_3}| \star \Psi^{\lambda_1}, |I \star \Psi^{\lambda_2}| \star \Psi^{\lambda_1}\right],
\end{equation}
where $\langle \cdot \rangle$ corresponds to the mean over the sphere, and where covariances are defined as $\text{Cov}[XY] = \langle X Y^*\rangle - \langle X \rangle \langle Y^* \rangle$ for two complex fields $X$ and $Y$.

\section{Generative model of the large scale structures}
\label{sec:gen_model}

\subsection{Maximum entropy generative model}

We build generative models under scattering covariance constraints. These are maximum entropy microcanonical models, which are approximately sampled by gradient descent~\cite{bruna2019multiscale}. Such models are constructed from statistics $\Phi$ estimated from a target field $x_t$; in this paper the target field is a single full-sky map. The associated microcanonical set $\Omega_\varepsilon$ of width $\varepsilon$ is
\begin{equation}
    \Omega_\varepsilon = \left\{ x : ||\Phi(x) - \Phi(x_t)||^2 < \varepsilon \right\},
\end{equation}
where $||.||$ is the Euclidean norm. The microcanonical maximum entropy model is the model of maximal entropy defined over $\Omega_\varepsilon$, which has an uniform distribution over this set.

In this paper, we approximate this sampling with a gradient descent approach, which consists in transporting a higher entropy Gaussian white noise distribution into a distribution supported in $\Omega_\varepsilon$. In practice, each new sample is obtained by first drawing a white noise realization, and then performing a gradient descent in pixel space using a loss
\begin{equation}
     \mathcal{L}(x) = ||\Phi(x) - \Phi(x_t)||^2.
\end{equation} 
The typical width $\varepsilon$ of the microcanonical ensemble is then fixed by the number of iterations used in the gradient descent. In practice we find that $\sim 100$ iterations is typically sufficient, in which case an image at $\texttt{nside} = 128$ may be generated in $\sim4$ seconds.

In our case, the summary statistics $\Phi(x)$ that we consider are the mean over pixels $\langle x \rangle$, its variance $\Var(x)$ and the SC statistics.

\subsection{Validation of the generative models}
\label{ss:results}
We compare the generated fields with the target one. We first do a visual comparison of the maps to asses the quality of the spatial texture reproduction. We then compute various standard statistics, such as the probability density function (PDF) and the angular power spectrum, in order to quantitatively evaluate our generative model. As we have drawn 50~samples of the microcanonical ensemble, we compute the mean and the standard deviation of these statistics over those 50~realisations.
We also propose a comparison with samples from a Gaussian model built from the power spectrum of the target field. This allows us to quantify the contribution of SC statistics compared to purely Gaussian statistics.

\begin{figure}[ht!]
    \centering 
    \includegraphics[width=1\linewidth]{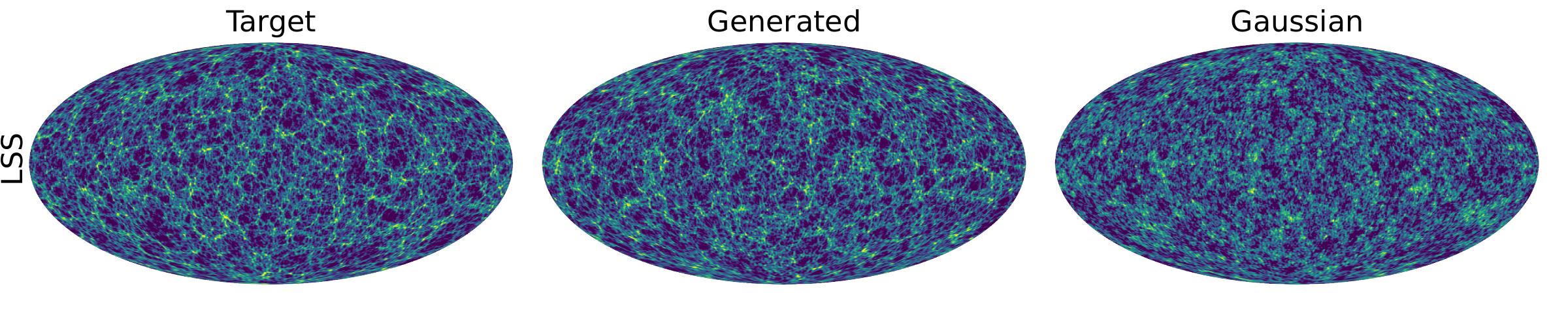}
    \caption{The original target field, a generated map and a Gaussian realisation. We plot the logarithm of the field and color bars are identical for all maps.}
    \label{fig:maps}
\end{figure}

In Fig.~\ref{fig:maps} we show the full-sky maps. We plot the logarithm of the maps in order to better see the textures. The generated map appears to be visually very similar to the target one, which clearly shows that the SC statistics capture an important part of the non-Gaussian texture of the field. On the contrary, the structures are not reproduced in the Gaussian realisation. 

\begin{figure*}
    \centering
    \includegraphics[width=0.32\linewidth]{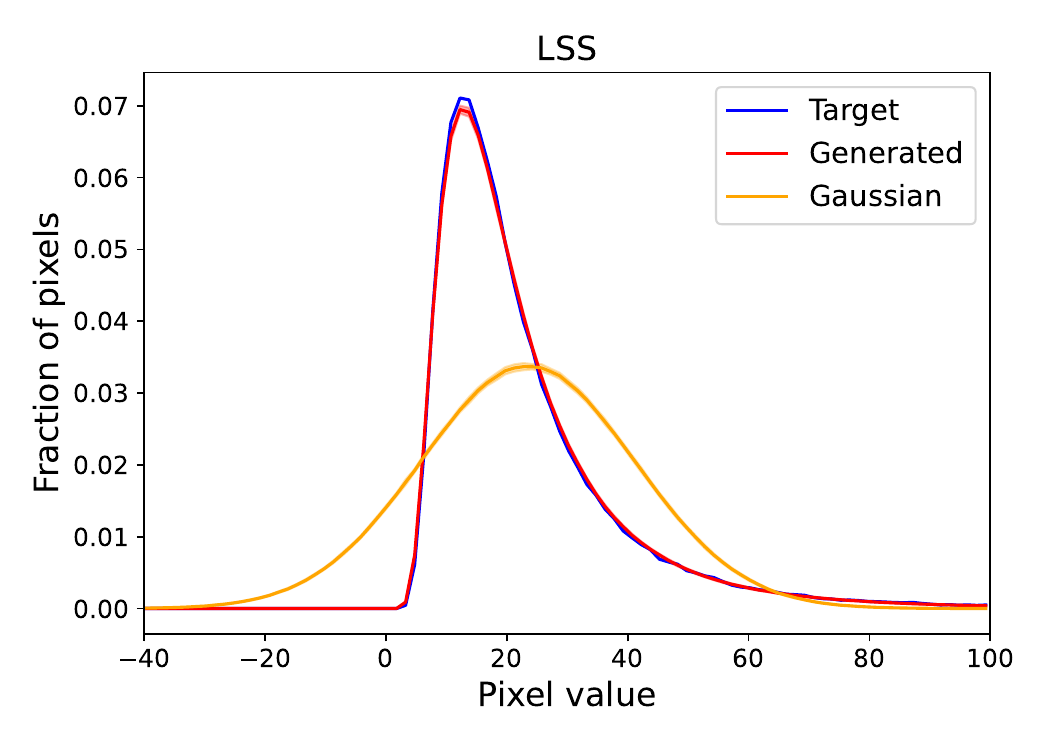}   
    \includegraphics[width=0.32\linewidth]{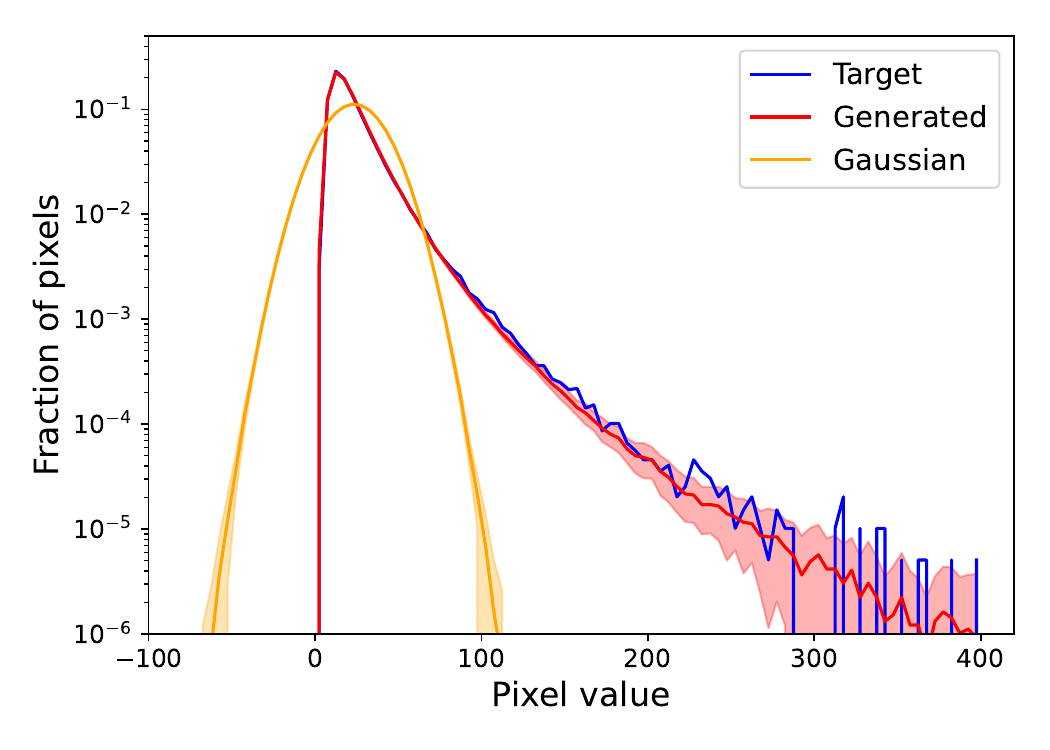}  
    \includegraphics[width=0.32\linewidth]{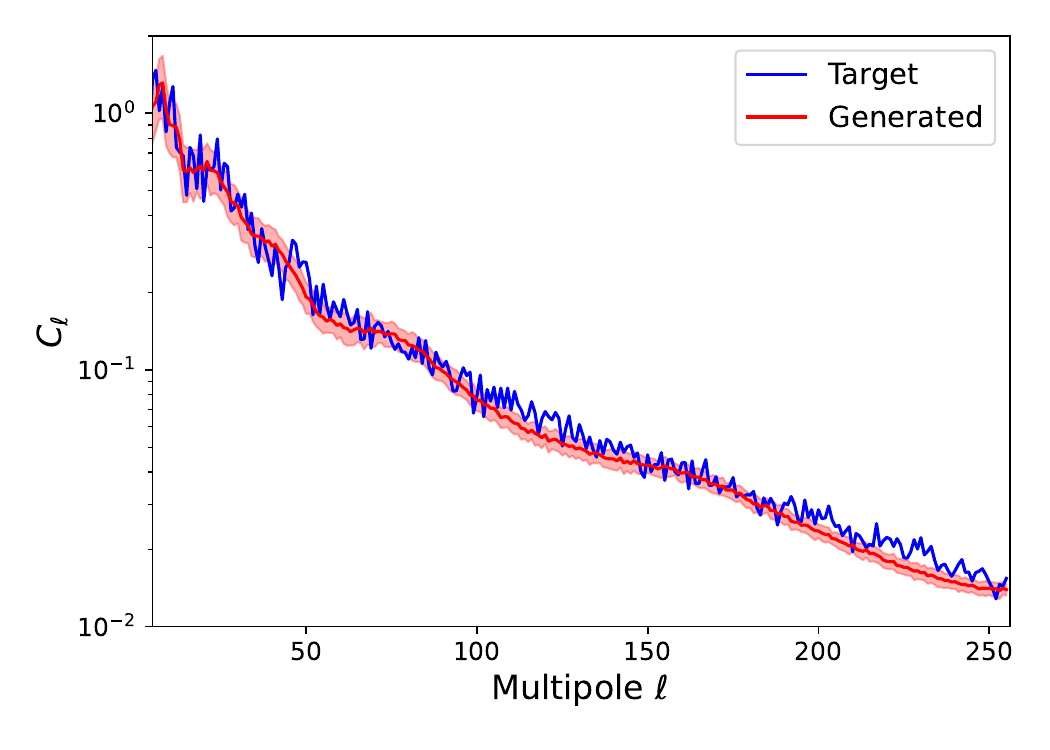} 
    \caption{PDF, both with a linear (left) and a logarithmic (middle) y-axis scaling, and angular power spectrum for the LSS field. The target is shown in blue, the generated fields in red, and the Gaussian realisations in yellow. We plot the mean (solid line) and the standard deviation (shadow envelope) over 50~realisations.}
    \label{fig:pdf_ps}
\end{figure*}

The PDF is shown in Fig.~\ref{fig:pdf_ps}, both with a linear (left) and a logarithmic (middle) y-axis scaling in order to better exhibit the tails of the distributions on several orders of magnitude. The target field is shown in blue, the generated ones in red and the Gaussian realisations in yellow. The comparison with the Gaussian PDF allows us to better see the non symmetric shape, which is characteristic of non-Gaussian features. As we can see, the PDF is well reproduced up to five orders of magnitude. 

The angular power spectrum is shown in Fig.~\ref{fig:pdf_ps} (right) and it is well reproduced over all scales. However, small oscillations around the target can be seen in the generated power spectra. These are residual features related to the frequency bands of the wavelets, which illustrate the trade-off between the quality of reproduction we want to achieve and the number of filters we use; i.e. the computational efficiency of our generative model. In Mousset et al. (2024)~\cite{mousset2024}, another validation using the Minkowski functionals, which are standard non-Gaussian statistics, is made, showing that they are also very well reproduced.

\section{Conclusions}
The main result of this work is the extension of state-of-the-art ST for generative modelling to spherical fields. We have worked with the last generation of ST statistics, named scattering covariances, which were previously introduced for one dimension and two dimensions planar fields. They have the advantage of relying only on successive wavelet transforms and modulus, as well as on covariances, and do not require any translations. We have also used state-of-the-art directional convolutions on the sphere~\cite{S2FFT_2023}, computed in spherical harmonic space. 

These developments allow us to build generative models of full sky spherical fields without the need for large training datasets. In fact, our method holds even in the limit of a single data realisation. In this proceeding, we have shown the result for the LSS weak lensing field for which they performed extremely well. In fact, the performance of those generative models were validated quantitatively on different fields and the diversity in terms of structures between the maps shows the impressive ability of SC to comprehensively characterize very different non-Gaussian textures~\cite{mousset2024}.

This work introduces a new powerful innovative approach for spherical data, and it opens interesting perspectives for astrophysical applications. In particular, we plan to use it for the study and the modeling of CMB astrophysical foregrounds. The first goal will be to have a tool to produce multiple realisations of the different astrophysical components. Then, ST could play a role in component separation, relying both on recently developed ST-based statistical component separations approaches, e.g., Auclair et al. (2024)~\cite{auclair2024}, as well as investigating how classical component separation methods could benefit from ST, using the non-Gaussianities as an additional lever arm to disentangle different components.



\end{document}